\documentstyle[12pt,epsfig,amssymb]{article}

\renewcommand{\theequation}{\mbox{$\thesection . \arabic{equation}$}}

\begin{document}
\newcommand{\beq}{\begin{equation}}
\newcommand{\eeq}{\end{equation}}
\newcommand{\barr}{\begin{eqnarray}}
\newcommand{\earr}{\end{eqnarray}}

\newcommand{\andy}[1]{ }

\def\rz{\mbox{\boldmath $y$}_0}
\def\a{\alpha_0} \def\da{\delta\alpha}
\def\Da{D_\alpha}
\def\h{\widehat}
\def\t{\widetilde}
\def\cH{{\cal H}}
\def\ud{\uparrow\downarrow}
\def\uu{\uparrow}
\def\dd{\downarrow}
\def\u{\uparrow}
\def\d{\downarrow}
\def\b1{{\bf 1}}

\def\coltwovector#1#2{\left({#1\atop#2}\right)}
\def\upp{\coltwovector10}    \def\downn{\coltwovector01}
\def\bra#1{\langle #1 |}
\def\ket#1{| #1 \rangle}


\def\ask{\marginpar{?? ask:  \hfill}}
\def\fin{\marginpar{fill in ... \hfill}}
\def\note{\marginpar{note \hfill}}
\def\check{\marginpar{check \hfill}}
\def\discuss{\marginpar{discuss \hfill}}


\begin{titlepage}
\begin{flushright}
\today \\
BA-TH/00-382\\
\end{flushright}
\vspace{.5cm}
\begin{center}
{\LARGE Quantum Zeno Dynamics

}

\quad

{\large P. FACCHI,$^{(1)}$ V. GORINI,$^{(2)}$ G. MARMO,$^{(3)}$ S.
PASCAZIO$^{(1)}$ and E.C.G. SUDARSHAN$^{(4)}$\\
           \quad    \\

        $^{(1)}$Dipartimento di Fisica, Universit\`a di Bari \\
       and Istituto Nazionale di Fisica Nucleare, Sezione di Bari \\
 I-70126  Bari, Italy \\
$^{(2)}$ Dipartimento di Scienze Chimiche, Fisiche e Matematiche \\
Universit\`a dell'Insubria, 22100 Como, Italy \\
       $^{(3)}$Dipartimento di Fisica, Universit\`a di Napoli \\
 Napoli, Italy \\
         $^{(4)}$Physics Department, University of Texas at Austin \\
 Texas 78712, USA \\

}

 \vspace*{.5cm} PACS: 03.65.Bz; 02.20.Mp; 02.30.Tb
\vspace*{.5cm}

{\small\bf Abstract}\\ \end{center}

{\small The evolution of a quantum system undergoing very frequent
measurements takes place in a subspace of the total Hilbert space
(quantum Zeno effect). The dynamical properties of this evolution are
investigated and several examples are considered.

}






\end{titlepage}

\newpage

\setcounter{equation}{0}
\section{Introduction }
 \label{sec-introd}
 \andy{intro}

A quantum system, prepared in a state that does not belong to an
eigenvalue of its total Hamiltonian, starts to evolve quadratically
in time \cite{Beskow,Misra}. This behavior leads to the so-called
quantum Zeno phenomenon: by performing frequent measurements on the
system, in order to check whether it is still in its initial state,
one can ``slow down" its temporal evolution (hindering transitions
to states different from the initial one) \cite{strev}.

This curious feature of the quantal evolution has recently attracted
much attention in the physics community. This is mainly due to a nice
idea put forward by Cook \cite{Cook}, who proposed to check this
effect on a two-level system, and to a related experimental test
\cite{Itano}, that motivated an interesting discussion \cite{Itanodiscuss}.
In turn, this has led to new proposals and experiments
\cite{PNBR,inn}. However, it should
be emphasized that these studies do not deal with {\em bona fide}
unstable systems, following (approximately) exponential laws, as in
the original proposals \cite{Beskow,Misra}. The presence of a
non-exponential decay at short times has been detected only recently
\cite{Wilkinson}.

The aim of the present paper is to investigate an interesting (and
often overlooked) feature of what we might call a quantum Zeno
dynamics. We shall see that a series of ``measurements" (von
Neumann's projections \cite{von}) does not necessarily hinder the
evolution of the quantum system. On the contrary, the system can
evolve away from its initial state, provided it remains in the
subspace defined by the ``measurement" itself. This interesting
feature is readily understandable in terms of a theorem proved by
Misra and Sudarshan (MS) \cite{Misra}, but it seems to us that it is
worth clarifying it further by analyzing some interesting examples.

\setcounter{equation}{0}
\section{Misra and Sudarshan's theorem }
\label{sec-MisSud}
\andy{MisSud}

Consider a quantum system Q, whose states belong to the Hilbert space
${\cal H}$ and whose evolution is described by the unitary operator
$U(t)=\exp(-iHt)$, where $H$ is a time-independent semi-bounded
Hamiltonian. Let $E$ be a projection operator that does not commute
with the Hamiltonian, $[E,H]\neq 0$, and $E{\cal H}E={\cal H}_E$ the
subspace spanned by its eigenstates. The initial density matrix
$\rho_0$ of system Q is taken to belong to ${\cal H}_E$. If Q is let
to follow its ``undisturbed" evolution, under the action of the
Hamiltonian $H$ (i.e., no measurements are performed in order to get
information about its quantum state), the final state at time $T$
reads
\andy{noproie}
\beq
\rho (T) = U(T) \rho_0 U^\dagger (T)
  \label{eq:noproie}
\eeq
and the probability that the system is still in ${\cal H}_E$ at time $T$
 is
\andy{stillun}
\beq
P(T) = \mbox{Tr} \left[ U(T) \rho_0 U^\dagger(T) E \right] .
\label{eq:stillun}
\eeq
We call this a ``survival probability:" it is in general smaller than
1, since the Hamiltonian $H$ induces transitions out of ${\cal H}_E$.
We shall say that the quantum system has ``survived" if it is found
to be in ${\cal H}_E$ by means of a suitable measurement process
\cite{MScomment}. We stress that we do not distinguish between one-
and many-dimensional projections: in the examples to be considered in
this note, $E$ will be infinite-dimensional.

Assume that we perform a measurement at time $t$, in order to check
whether Q has survived. Such a measurement is formally represented by
the projection operator $E$. By definition,
\andy{inprep}
\beq
\rho_0 = E \rho_0 E , \qquad \mbox{Tr} [ \rho_0 E ] = 1 .
\label{eq:inprep}
\eeq
After the measurement, the state of Q changes into
\andy{proie}
\beq
\rho_0 \rightarrow \rho(t) = E U(t) \rho_0 U^\dagger(t) E,
\label{eq:proie}
\eeq
with probability
\andy{probini}
\barr
P(t) &=& \mbox{Tr} \left[ U(t) \rho_0 U^\dagger(t) E \right] =
\mbox{Tr} \left[E U(t)E \rho_0 E U^\dagger(t) E \right] \nonumber
\\
 &=& \mbox{Tr} \left[V(t) \rho_0 V^\dagger(t) \right].
  \quad \qquad (V(t) \equiv E U(t)E)
\label{eq:probini}
\earr
This is the probability that the system has ``survived" in ${\cal
H}_E$. There is, of course, a probability $1-P$ that the system has
not survived (i.e., it has made a transition outside ${\cal H}_E$)
and its state has changed into $\rho^\prime(t) = (1-E) U(t)
\rho_0 U^\dagger(t) (1-E)$. The states $\rho$ and $\rho'$ together
make up a block diagonal matrix: The initial density matrix is
reduced to a mixture and any possibility of interference between
``survived" and ``not survived" states is destroyed (complete
decoherence).

We shall concentrate henceforth our attention on the measurement
outcome (\ref{eq:proie})-(\ref{eq:probini}). We observe that the
evolution just described is time-translation invariant and the
dynamics is {\em not} reversible (not only not time-reversal
invariant).

The above is the Copenhagen interpretation: the measurement is
considered to be instantaneous. The ``quantum Zeno paradox"
\cite{Misra} is the following. We prepare Q in the initial state
$\rho_0$ at time 0 and perform a series of $E$-observations at times
$t_j=jT/N \; (j=1,
\cdots, N)$. The state of Q after the above-mentioned $N$
measurements reads
\andy{Nproie}
\beq
\rho^{(N)}(T) = V_N(T) \rho_0 V_N^\dagger(T) , \qquad
    V_N(T) \equiv [ E U(T/N) E ]^N
\label{eq:Nproie}
\eeq
and the probability to find the system in ${\cal H}_E$ (``survival
probability") is given by
\andy{probNob}
\beq
P^{(N)}(T) = \mbox{Tr} \left[ V_N(T) \rho_0 V_N^\dagger(T) \right].
\label{eq:probNob}
\eeq
Equations (\ref{eq:Nproie})-(\ref{eq:probNob}) display the ``quantum
Zeno effect:" repeated observations in succession modify the dynamics
of the quantum system; under general conditions, if $N$ is
sufficiently large, all transitions outside ${\cal H}_E$ are
inhibited. Notice again that the dynamics
(\ref{eq:Nproie})-(\ref{eq:probNob}) is not reversible.

In order to consider the $N \rightarrow \infty$ limit (``continuous
observation"), one needs some mathematical requirements: assume
that the limit
\andy{slim}
\beq
{\cal V} (T) \equiv \lim_{N \rightarrow \infty} V_N(T)
  \label{eq:slim}
\eeq
exists in the strong sense. The final state of Q is then
\andy{infproie}
\beq
\rho (T) = {\cal V}(T) \rho_0 {\cal V}^\dagger (T)
  \label{eq:infproie}
\eeq
and the probability to find the system in ${\cal H}_E$ is
\andy{probinfob}
\beq
{\cal P} (T) \equiv \lim_{N \rightarrow \infty} P^{(N)}(T)
   = \mbox{Tr} \left[ {\cal V}(T) \rho_0 {\cal V}^\dagger(T) \right].
\label{eq:probinfob}
\eeq
One should carefully notice that nothing is said about the final
state $\rho (T)$, which depends on the characteristics of the model
investigated and on the {\em very measurement performed} (i.e.\ on
the projection operator $E$, by means of which $V_N$ is defined). By
assuming the strong continuity of ${\cal V}(t)$
\andy{phgr}
\beq
\lim_{t \rightarrow 0^+} {\cal V}(t) = E,
\label{eq:phgr}
\eeq
one can prove that under general conditions the operators
\andy{semigr}
\beq
{\cal V}(T) \quad \mbox{exist for all real $T$ and form a
semigroup.}
\label{eq:semigr}
\eeq
Moreover, by time-reversal invariance
\andy{VVdag}
\beq
{\cal V}^\dagger (T) = {\cal V}(-T),
\label{eq:VVdag}
\eeq
so that ${\cal V}^\dagger (T) {\cal V}(T) =E$. This implies, by
(\ref{eq:inprep}), that
\andy{probinfu}
\beq
{\cal P}(T)=\mbox{Tr}\left[\rho_0{\cal V}^\dagger(T){\cal V}(T)\right]
= \mbox{Tr} \left[ \rho_0 E \right] = 1 .
\label{eq:probinfu}
\eeq
If the particle is ``continuously" observed, in order to check
whether it has survived inside ${\cal H}_E$, it will never make a
transition to  ${\cal H}_E^\perp$. This was named ``quantum Zeno
paradox" \cite{Misra}. The expression ``quantum Zeno effect" seems
more appropriate, nowadays.

Two important remarks are now in order: first, it is not clear
whether the dynamics in the $N \to \infty$ limit is time reversible.
Although one ends up, in general, with a semigroup, there are
concrete elements of reversibility in the above equations. Second,
the theorem just summarized {\em does not} state that the system {\em
remains} in its initial state, after the series of very frequent
measurements. Rather, the system is left in the subspace ${\cal
H}_E$, instead of evolving ``naturally" in the total Hilbert space
${\cal H}$. This subtle point, implied by Eqs.\
(\ref{eq:infproie})-(\ref{eq:probinfu}), is not duely stressed in the
literature \cite{MNPRY}.

Incidentally, we stress that there is a conceptual gap between
Eqs.\ (\ref{eq:probNob}) and (\ref{eq:probinfob}): to perform an
experiment with $N$ finite is only a practical problem, from the
physical point of view. On the other hand, the $N \rightarrow
\infty$ case is physically unattainable, and is rather to be
regarded as a mathematical limit (although a very interesting one).
In this paper, we shall not be concerned with this problem
(investigated in \cite{NNPR}; see also \cite{BerryKlein}, where an
interesting perspective is advocated) and shall consider the $N
\to \infty$ limit for simplicity. This will make the analysis more
transparent.

\setcounter{equation}{0}
\section{Evolution in the ``Zeno" subspace}
\andy{compact}
\label{sec-compact}

We start off by looking at some explicit examples. Consider a free
particle of mass $m$ on the real line. The Hamiltonian and the
corresponding evolution operator are
\andy{freeHam}
\beq
H=\frac{p^2}{2m},\qquad U(t)=\exp(-i t H ).
\label{eq:freeHam}
\eeq
Observe that $H$ is a positive-definite self-adjoint operator in
$L^2(\mathbb{R})$ and $U(t)$ is unitary. We shall study the quantum
Zeno effect when the system undergoes a measurement defined by the
projector
\andy{proja}
\beq
E_A=\int dx\;\chi_A(x) \ket{x}\bra{x}, \label{eq:proja}
\eeq
where $\chi_A$ is the characteristic function
\andy{chara}
\beq
\chi_A(x)=\left\{
  \begin{array}{l}
    1 \quad \mbox{for } x \in A \subset \mathbb{R} \\
    0 \quad \mbox{otherwise}
  \end{array}\right.
\label{eq:chara}
\eeq
and $A$ an interval of $\mathbb{R}$. In a few words, we check whether
a particle, initially prepared in a state with support in $A$ and
free to move on the real line, is still found in $A$ at a later time
$T$. Our objective is to understand how the system evolves in the
``Zeno" subspace ${\cal H}_{E_A}=E_A{\cal H}E_A$. We call this a
``quantum Zeno dynamics with a nonholonomic constraint."

We shall work with the Euclidean Feynman integral. Let the particle
be initially ($t=0$) at position $y \in E_A$. The propagator at time
$t=T/N$, when a measurement is carried out, reads
\beq
G(x,t;y) \equiv \bra{x} E_A\, U(t)
\ket{y}=\chi_A(x)\bra{x}U(t)\ket{y}.
\eeq
For imaginary time $t=-i\tau$, we get the Green function of the
heat equation
\barr
\bra{x}U(-i\tau)\ket{y}&=&\bra{x}\exp(-\tau H)\ket{y}
=\int dp\; \bra{x} p\rangle e^{-\tau p^2/2m} \langle p \ket{y}
\nonumber\\
&=&\int \frac{dp}{2\pi}\; e^{-\tau p^2/2m+i p(x-y)}
=\sqrt{\frac{m}{2\pi\tau}}\exp\left[-\frac{m(x-y)^2}{2\tau}\right],
\earr
so that the Euclidean propagator for a single ``step" reads
\andy{euclisingle}
\beq
W(x,\tau;y) \equiv G(x,-i\tau;y)=\chi_A(x)
\sqrt{\frac{m}{2\pi\tau}}\exp\left[-\frac{m(x-y)^2}{2\tau}\right].
\label{eq:euclisingle}
\eeq
The evolution operator after $N$ measurements, see (\ref{eq:Nproie}),
can be written as
\beq
 V_N(T) \equiv [ E_A\, U(T/N)]^N E_A
\eeq
and the resulting propagator is
\beq
G_N(x_{\rm f},T;x_{\rm i})=\bra{x_{\rm f}} V_N(T)\ket{x_{\rm i}}.
\eeq
For imaginary ${\cal T}=i T$ this becomes
\andy{Wiener}
\barr
W_{N}(x_{\rm f},{\cal T};x_{\rm i}) & \equiv & G_N(x_{\rm f},-i {\cal
T} ;x_{\rm i})
\nonumber
\\ &=&\int dx_1\cdots dx_{N-1} W\left(x_{\rm
f},\tau;x_{N-1}\right)\cdots W\left(x_1,\tau;x_{\rm i}\right)
\chi_A(x_i),
\label{eq:Wiener}
\earr
whose relation with Wiener integration is manifest. Notice that if we
could drop the characteristic function $\chi_A$ in the propagator
(\ref{eq:euclisingle}), then (\ref{eq:Wiener}) would be a sequence of
nested Gaussian integrals, that could be evaluated exactly for every
$N$ by applying Feynman's recipe
\cite{FeynmanHibbs}. In (\ref{eq:Wiener}) the characteristic
functions restrict at every step the set of possible paths, modifying
the structure of the functional integral \cite{Yamada}. Let us
therefore try to reduce the integral (\ref{eq:Wiener}) to a Gaussian
form. To this end we apply a trick that is often used when one
endeavours to relate probability and potential theory
\cite{Schulman}. We first rewrite the characteristic function in
terms of a potential, which is infinite outside $A$ \cite{foot}, so
that the Brownian paths of the Wiener process (\ref{eq:Wiener}) can
never leak out of $A$:
\andy{potential}
\beq
\chi_A(x)=\exp\left(-\tau V_A(x)\right),\qquad
\mbox{with}\qquad
V_A(x)=\left\{
  \begin{array}{l}
    0 \quad \mbox{for } x \in A \\
    +\infty \quad \mbox{otherwise}
  \end{array}\right. ,
\label{eq:potential}
\eeq
Hence, by using (\ref{eq:potential}), the Euclidean one-step
propagator (\ref{eq:euclisingle}) becomes
\barr
W(x,\tau;y)=
\sqrt{\frac{m}{2\pi\tau}}
\exp\left[-\frac{m(x-y)^2}{2\tau}-\tau V_A(x)\right]
=\bra{x}e^{-\tau V_A}e^{-\tau H} \ket{y}
\earr
and returning to real time
\beq
G(x,t;y)=W(x,it;y)=\bra{x}e^{-it V_A}e^{-it H}\ket{y}.
\eeq
Consider now the limit of continuous observation $N\to\infty$. The
limiting propagator reads
\beq
{\cal G}(x_{\rm f},T ;x_{\rm i})=\lim_{N\to\infty}G_N(x_{\rm
f},T;x_{\rm i})=\lim_{N\to\infty}\bra{x_{\rm f}}
\left(e^{-iT V_A/N}e^{-iT H/N} \right)^N E_A\ket{x_{\rm i}} ,
\eeq
which, by using the Trotter product formula, yields
\beq
{\cal G}(x_{\rm f},T ;x_{\rm i}) =
\bra{x_{\rm f}}e^{-iT(H+V_A)} E_A\ket{x_{\rm i}}
=\bra{x_{\rm f}} {\cal V}(T) \ket{x_{\rm i}},
\eeq
where the evolution operator is
\andy{VT,HZ}
\barr
& & {\cal V}(T)=\exp(-iT H_{\rm Z})\, E_A,
\label{eq:VT} \\
& & \qquad\mbox{with}\qquad H_{\rm Z}\equiv\frac{p^2}{2m}+V_A(x).
\label{eq:HZ}
\earr
The above formula is of general validity: the dynamics {\em within}
the Zeno subspace ${\cal H}_{E_A}$ is governed by the operators
(\ref{eq:VT})-(\ref{eq:HZ}).

It is worth stressing that the previous calculation only makes use of
the properties of the kinetic energy operator $p^2$: we have {\em
not} considered the momentum operator $p$. It goes without saying
that $p$ can be symmetric, maximally symmetric or self-adjoint,
according to the structure of $A$ and the boundary conditions. This
will be thoroughly discussed in the following. However, we emphasize
that any requirement on $p$ would be a {\em physical} requirement:
the mathematical properties of the ``Zeno" evolution {\em only}
involve the Hamiltonian (which is defined in terms of the kinetic
energy).

Before we proceed further, let us look at two particular cases:
\andy{A1,A2}
\barr
A_1&=&[0,1],
\label{eq:A1} \\
A_2&=&[0,+\infty).
\label{eq:A2}
\earr
In the first case, the free Hamiltonian
\andy{HZ0}
\beq
H_{\rm Z}^0=\frac{-\partial^2}{2m}    \quad
\left(\partial\equiv \frac{d}{dx} \right)
\label{eq:HZ0}
\eeq
is a self-adjoint operator on the space
\andy{AC201}
\beq
D_{[0,1]}(H_{\rm Z}^0) = \left\{\phi \in \mbox{AC}^2[0,1] |
\phi(0)=\phi(1)=0 \right\},
\label{eq:AC201}
\eeq
where AC$^2[S]$ is the set of functions in $L^2[S]$ whose weak
derivatives are in AC$[S]$. (AC$[S]$ is the set of absolutely
continuous functions whose weak derivatives are in $L^2[S]$.) Notice
that these are the ``correct" boundary condition for the potential
(\ref{eq:potential}). For this reason, the evolution operators ${\cal
V}(T)$ in (\ref{eq:VT}) form a one-parameter {\em group}. We notice,
incidentally, that MS's mathematical hypotheses (\ref{eq:slim}) and
(\ref{eq:phgr}) are satisfied and acquire in this example an
appealing physical meaning. We also stress that the theorem
(\ref{eq:semigr}) appears in this case too restrictive: indeed the
operators ${\cal V}(T)$ form a group and not simply a semigroup. One
might say that in the example considered, the quantum Zeno effect
(engendered by the projection operators) automatically yields the
``natural" dynamics in the Zeno subspace, with the correct boundary
conditions for the ``new" Hamiltonian $H_{\rm Z}$. This is an
interesting observation in itself. We also notice that in this
example the momentum operator $-i\partial$ is symmetric, but {\em
not} self-adjoint: its deficiency indices in  (\ref{eq:AC201}) are
(1,1). Therefore, a self-adjoint extension of $-i\partial$ is
possible. It is important to stress that the Hamiltonian $H_{\rm Z}$
is self-adjoint because it involves only $\partial^2$ [which is
self-adjoint in (\ref{eq:AC201})]. There is here an interesting
classical analogy: when a classical particle elastically bounces
between two rigid walls, any trajectory is characterized by a
definite value of energy ($p^2/2m$), although momentum changes
periodically between $\pm p$. This is reflected in the symmetry
(rather than self-adjointness) of the quantum mechanical $p$
operator.

Let us now look at the example $A_2$ in (\ref{eq:A2}). The free
Hamiltonian (\ref{eq:HZ0}) is self-adjoint on the space
\andy{AC20inf}
\beq
D_{[0,\infty)}(H_{\rm Z}^0) = \left\{\phi \in \mbox{AC}^2[0,\infty)
|
\phi(0)=0 \right\},
\label{eq:AC20inf}
\eeq
Once again, this is just the ``correct" boundary condition for the
potential (\ref{eq:potential}), so that the evolution operators
${\cal V}(T)$ form a one-parameter group. One can draw the same
conclusions as in the previous example. There is only one difference:
the momentum operator $-i\partial$ is again symmetric, but its
deficiency indices are (0,1). This is irrelevant as far as one's
attention is restricted to the Hamiltonian and the Zeno dynamics;
however, if one is motivated (on physical grounds) to consider the
properties of momentum, the best one can do in this case is to obtain
the most appropriate maximally symmetric momentum operator. (We
wonder whether this has spin-offs at a fundamental quantum mechanical
level.)

\setcounter{equation}{0}
\section{The problem of the lower-boundedness of the Hamiltonian }
 \label{sec-lb}
 \andy{lb}

Let us consider now the model Hamiltonian $H^0_{\rm Z}=p=-i\partial$
in $A_1=[0,1]$, describing an ultrarelativistic particle in an
interval. The mathematical features of this example are very
interesting and deserve careful investigation. A similar example was
considered in \cite{Misra}, although in a different perspective. The
Zeno dynamics yields
\andy{VT1,HZ1}
\barr
& & {\cal V}(T)=\exp(-iT H_{\rm Z})\, E_{A_1},
\label{eq:VT1} \\
& & \qquad\mbox{with}\qquad H_{\rm Z}\equiv p+V_{A_1}(x),
\label{eq:HZ1}
\earr
where $V_A$ is defined in (\ref{eq:potential}). The ``natural"
boundary conditions imposed by the Zeno dynamics are
\andy{Dnat}
\beq
D_{\rm Z}(p)= \left\{\phi \in \mbox{AC}[0,1] | \phi(0)=0=\phi(1)
\right\}.
\label{eq:Dnat}
\eeq
In this domain the Hamiltonian $p$ is symmetric but {\it not}
self-adjoint: its deficiency indices are $(1,1)$. Therefore, by
Stone's theorem, the Zeno dynamics is not governed by a group and is
certainly not time-reversal invariant. More to this, this Hamiltonian
is not lower bounded and therefore violates one of the premises of
the MS theorem. In order to understand what happens during a Zeno
dynamics, look at the first row in Figure~1, where an arbitrary wave
packet evolves under the action of the free Hamiltonian $p$
(incidentally, notice that the wave packet does not disperse, due to
the form of the Hamiltonian). The probability of ``surviving" inside
$A_1$ decreases with time: in other words, even though a
``continuous" measurement is performed, in  order to check whether
the particle is outside $A_1$, the particle {\em does} leak out of
$A_1$ and {\em no} quantum Zeno effect takes place.

Let us now assume, on physical grounds, the validity of periodic
boundary conditions:
 \andy{Dper}
\beq
D^{\alpha}(p)= \left\{\phi \in \mbox{AC}[0,1] |
\phi(0)=\phi(1) e^{i\alpha} \right\},
\label{eq:Dper}
\eeq
where the phase $\alpha$ determines the specific self-adjoint
extension. Notice that this is a physical requirement: it is not a
consequence of the Zeno dynamics. The Hamiltonian is now self-adjoint
and the dynamics is governed by a unitary group (Stone's theorem).
Obviously, the physical picture given by this self-adjoint extension
is completely different from the previous case. See the second row in
Figure~1: a quantum Zeno effect takes place.
\begin{figure}
\epsfig{file=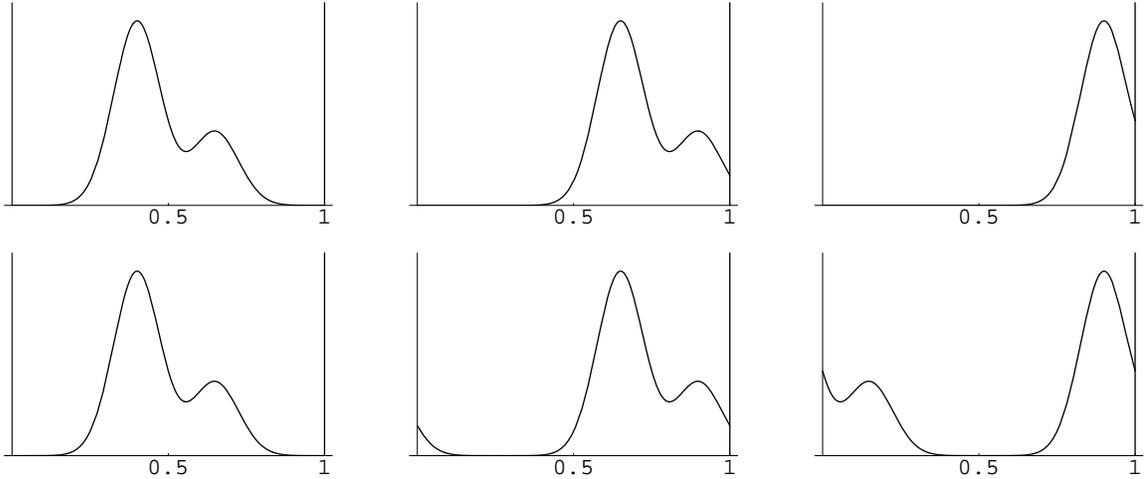,width=\textwidth}
\caption{``Natural" (Zeno) vs. periodic (self-adjoint) boundary conditions
for the Hamiltonian $H=p$. Above: quantum evolution of a wave packet
of arbitrary shape with the boundary conditions (\ref{eq:Dnat}),
``naturally" arising in a Zeno dynamics: there is no quantum Zeno
effect (increasing time from left to right). Below: evolution of the
same wave packet with the additional requirement that the unbounded
Hamiltonian operator $p$ be self-adjoint in $[0,1]$: a quantum Zeno
effect occurs (increasing time from left to right). }
\label{fig:compy}
\end{figure}

We also stress that the dependence of the Hamiltonian on the $p$
operator is not a sufficient condition to yield the behavior
described above. In order to clarify this point, let us consider an
additional example. Let (we set $m=1/2$)
\andy{Hnew}
\beq
H=p^2+p \quad \Longleftrightarrow \quad H_{\rm Z}=p^2+p+V_A(x).
\label{eq:Hnew}
\eeq
We first observe that $H$ is lower bounded [$p^2+p=(p+1/2)^2-1/4$;
notice also that this Hamiltonian can be tranformed into the usual
form by adding a phase $x/2$ to the wave function.] Consider again
the quantum Zeno dynamics on the sets $A_1$ and $A_2$. Since this is
not a classical textbook example, we explicitly derive the
deficiencies. In the first case ($A_1$) one gets
\andy{defH}
\barr
(H_{\rm Z}\phi,\psi) - (\phi,H^*_{\rm Z}\psi) &=& -i
\overline{\phi(0)}
\psi(0) +
\overline{\phi^{\prime}(0)} \psi(0) - \overline{\phi(0)} \psi^\prime(0) \nonumber \\
& & + i \overline{\phi(1)} \psi(1) -
\overline{\phi^{\prime}(1)} \psi(1) + \overline{\phi(1)} \psi^\prime(1).
\label{eq:defH}
\earr
It is easy to check that $H_{\rm Z}$ is lower bounded and self
adjoint on the space (\ref{eq:AC201}). The Zeno evolution is
therefore unitary.

In the second case ($A_2$) one gets
\andy{defH1}
\beq
(H_{\rm Z}\phi,\psi) - (\phi,H_{\rm Z}^*\psi) = -i
\overline{\phi(0)}
\psi(0) + \overline{\phi^{\prime}(0)} \psi(0) - \overline{\phi(0)}
\psi^\prime(0).
\label{eq:defH1}
\eeq
It is straightforward to check that the Hamiltonian is lower bounded
and self-adjoint on the space (\ref{eq:AC20inf}). Once again, the
Zeno evolution is unitary.

\setcounter{equation}{0}
\section{Discussion}
\andy{disc}
\label{sec-disc}

One is led to the following question: is it possible to find an
example in which the Zeno dynamics is governed by a dynamical {\em
semigroup}? The answer to this question would be positive if one
could find a quantum Zeno dynamics yielding a symmetric, but not
self-adjoint, Hamiltonian operator. Indeed, in such a case, by
Stone's theorem one cannot have a group, and by MS's theorem one must
have a semigroup.

It would be incorrect to think that the model Hamiltonian
$H=p=-i\partial$ in $A_1=[0,1]$ (or even more in $A_2=[0,\infty)$)
provides us with the counterexample we seek. Indeed, such a
Hamiltonian is not a satisfactory example, because it violates one of
the premises of the MS theorem, that requires a lower-bounded
Hamiltonian from the outset (see beginning of Section 2).

We are unable, at the present stage, to give a clear-cut answer to
this problem. However, some comments are in order. If, for some
reason, the quantum Zeno dynamics yields a symmetric Hamiltonian
operator, the search for its self-adjoint extensions seems to us a
very important one, on physical grounds. Suppose then that one is
willing to consider a self-adjoint extension of the Zeno Hamiltonian
$H_{\rm Z}$. If this is the
 case, close inspection shows that a quantum Zeno dynamics always yields a
group, at least in the class of systems considered in this note.
Indeed, a theorem due to von Neumann, Stone and Friedrichs
\cite{ReedSimon} states that ``every semi-bounded symmetric
transformation $S$ can be extended to a semi-bounded self-adjoint
transformation $S'$ in such a way that $S'$ has the same (greatest
lower or least upper) bound as $S$." Therefore, if the Hamiltonian is
lower-bounded on the real line, as for instance in (\ref{eq:freeHam})
(one could even add a non-pathological potential to the kinetic
energy), the Zeno dynamics in an interval of $\mathbb{R}$ will also
be engendered by a lower-bounded Hamiltonian, like in (\ref{eq:HZ});
this would always admit a self-adjoint extension (due to the
above-mentioned theorem), which in turn would yield a {\em group} of
evolution operators. Therefore, in order to avoid the consequences of
von Neumann's theorem, the operators arising from the Zeno dynamics
must {\em not} be lower bounded. Only in such a case the Zeno
Hamiltonian might not admit self-adjoint extensions.

In conclusion, we have seen that in the situations considered in this
paper the quantum Zeno effect yields a {\em unitary} dynamics,
governed by groups, not by semigroups. We are therefore left with two
possible options: i) The MS theorem can be made stronger and the Zeno
dynamics is {\em always} governed by a group; ii) Different
projections, more general than (\ref{eq:proja})-(\ref{eq:chara}),
and/or different Hamiltonian operators may yield symmetric Zeno
Hamiltonian operators that are not self-adjoint (or, even more,
maximally symmetric operators with no self-adjoint extensions) and
therefore (due to the MS theorem) a {\em semi}group of evolution
operators.

 The answer to the above alternative would clarify whether
a quantum Zeno dynamics introduces some elements of irreversibility
in the evolution of a quantum system. This is an interesting open
problem.

\section*{Acknowledgments}
We thank I. Antoniou for interesting remarks.



\begin{thebibliography}{99}
\bibitem{MScomment} \andy{MScomment}
Misra and Sudarshan in \cite{Misra} considered an initial
``unstable" state $\rho_0$ and took $E$ to be the subspace of the
undecayed states. The expression ``survived" means in their case
``undecayed."
\bibitem{foot} \andy{foot} It is only at this stage
that we must require that $A$ is a well-behaved set like, for
example, an interval of $\mathbb{R}$ [see our starting point
(\ref{eq:chara})]. For instance, the potential (\ref{eq:potential})
would be physically meaningless for the Cantor set in $[0,1]$.
\end{thebibliography}
\end{document}